\documentclass[runningheads]{llncs}
\usepackage[T1]{fontenc}
\usepackage{graphicx}
\usepackage{cite}
\usepackage{amsmath,amssymb,amsfonts}
\usepackage{graphicx}
\usepackage{textcomp}
\usepackage{algorithm,algorithmic}
\usepackage{multicol}
\usepackage{multirow}
\usepackage{changepage,threeparttable}
\usepackage{makecell}
\usepackage{array}
    \newcolumntype{P}[1]{>{\centering\arraybackslash}p{#1}}
    \newcolumntype{M}[1]{>{\centering\arraybackslash}m{#1}}
\usepackage{booktabs, multirow}
\usepackage{soul}
\usepackage[T1]{fontenc}

\usepackage{amsmath,amsfonts}
\usepackage{algorithmic}
\usepackage{array}
\usepackage[caption=false,font=normalsize,labelfont=sf,textfont=sf]{subfig}
\usepackage{textcomp}
\usepackage{stfloats}
\usepackage{url}
\usepackage{verbatim}
\usepackage{graphicx}
\def\BibTeX{{\rm B\kern-.05em{\sc i\kern-.025em b}\kern-.08em
    T\kern-.1667em\lower.7ex\hbox{E}\kern-.125emX}}
\usepackage{balance}

\usepackage[colorlinks,urlcolor=blue,linkcolor=blue,citecolor=blue]{hyperref}

\usepackage{color,array}
\usepackage[utf8]{inputenc}
\usepackage{amsmath,amssymb,amsfonts}
\usepackage{algorithm,algorithmic}
\usepackage[mathscr]{euscript}

\usepackage{orcidlink}

\usepackage{comment}
\usepackage{hyperref}

\begin{document}
\title{An All-in-one Approach for Accelerated Cardiac MRI Reconstruction}
\titlerunning{All-in-one CMRI Reconstruction}
%
\author{}
\institute{}
\author{Kian Anvari Hamedani\inst{}\orcidlink{0009-0005-7416-1987} \and
Narges Razizadeh\inst{}\orcidlink{0009-0007-8985-9653} \and
Shahabedin Nabavi\inst{*}\orcidlink{0000-0001-7240-0239} \and
Mohsen Ebrahimi Moghaddam\inst{}\orcidlink{0000-0002-7391-508X}}
\authorrunning{K. Anavari Hamedani et al.}
%
\institute{Faculty of Computer Science and Engineering, Shahid Beheshti University, Tehran, Iran \\
\email{\{k.anvarihamedani,n.razizade\}@mail.sbu.ac.ir}\\
\email{\{s\_nabavi,m\_moghadam\}@sbu.ac.ir}\\
\href{https://github.com/kiananvari/CMRRecon}{https://github.com/kiananvari/CMRRecon}
}
\maketitle              
\begin{abstract}
Cardiovascular magnetic resonance (CMR) imaging is the gold standard for diagnosing several heart diseases due to its non-invasive nature and proper contrast. MR imaging is time-consuming because of signal acquisition and image formation issues. Prolonging the imaging process can result in the appearance of artefacts in the final image, which can affect the diagnosis. It is possible to speed up CMR imaging using image reconstruction based on deep learning. For this purpose, the high-quality clinical interpretable images can be reconstructed by acquiring highly undersampled k-space data, that is only partially filled, and using a deep learning model. In this study, we proposed a stepwise reconstruction approach based on the Patch-GAN structure for highly undersampled k-space data compatible with the multi-contrast nature, various anatomical views and trajectories of CMR imaging. The proposed approach was validated using the CMRxRecon2024 challenge dataset and outperformed previous studies. The structural similarity index measure (SSIM) values for the first and second tasks of the challenge are 99.07 and 97.99, respectively. This approach can accelerate CMR imaging to obtain high-quality images, more accurate diagnosis and a pleasant patient experience. 
\end{abstract}
\section{Introduction}
Cardiovascular magnetic resonance (CMR) imaging is a powerful and versatile tool in cardiology, offering detailed and comprehensive information on cardiac anatomy, function, and tissue characteristics \cite{arnold2020cardiovascular}. Its non-invasive nature, lack of ionizing radiation, and ability to provide quantitative and reproducible data make it an invaluable resource in diagnosing, treating, and managing a wide range of cardiac conditions \cite{vasquez2019clinical}. Despite its many advantages, CMR imaging faces challenges related to patient factors and technical limitations \cite{raman202230}. Patients are often required to hold their breath during image acquisition to minimize motion artefacts \cite{nabavi2023generalised, nabavi2024statistical}. This can be challenging for patients with respiratory issues or those who cannot hold their breath for prolonged periods. The bounded space of the MR scanner can induce claustrophobia in some patients, making it difficult for them to remain still and calm during the scan time \cite{enders2011reduction}. Thus, reducing acquisition time is one of the main objectives of researchers in this field.

Several reconstruction techniques, such as parallel imaging, compressed sensing, real-time imaging, optimized k-space sampling, AI-enhanced methods, and optimized imaging protocols, can significantly reduce scan times in CMR imaging. Traditional CMR image reconstruction relies on the inverse Fourier transform, which requires full k-space data for high-quality images. Parallel Imaging techniques like sensitivity encoding (SENSE) \cite{pruessmann1999sense} and generalized auto-calibrating partially parallel acquisitions (GRAPPA) \cite{griswold2002generalized} reduced scan time by under-sampling k-space and reconstructing images using coil sensitivity profiles. Compressed sensing (CS) \cite{lustig2007sparse} proposed a reconstruction method from under-sampled k-space data utilizing the sparsity of MR images in specified domains. Non-Cartesian k-space sampling \cite{block2007undersampled} and partial Fourier imaging \cite{bydder2005partial} can also be mentioned among the fast MRI reconstruction methods. In recent years, deep learning algorithms have been widely exploited for high-quality image reconstruction from sparsely sampled k-spaces, reducing extensive data acquisition \cite{lyu2024state}. Convolutional neural networks (CNNs) \cite{schlemper2017deep, xin2023fill, yiasemis2023deep}, generative adversarial networks (GANs) \cite{yang2017dagan}, recurrent neural networks (RNNs) \cite{qin2018convolutional} and hybrid approaches \cite{hammernik2018learning} have demonstrated superior performance than other methods.

The current study proposes a universal approach to multi-contrast CMR reconstruction with various acceleration factors and k-space trajectories in the form of an integrated model. The contributions of the study include the following:
\begin{itemize}
    \item An all-in-one approach is proposed based on the Patch-GAN architecture that reconstructs the subsampled k-space in a stepwise manner. 
    \item The curriculum learning approach is used in the model training process to achieve better performance.
    \item The combined loss function was defined by considering the physical properties of the k-space and features of the image domain. 
    \item The stepwise loss calculation was used to overcome the vanishing gradient problem.
\end{itemize}

\section{Materials and Methods}
The details of the datasets, the proposed approach, and the training and evaluation procedure are described in this section.

\subsection{Datasets}
The datasets used for training the proposed method include CMRxRecon2023 \cite{wang2024cmrxrecon} and CMRxRecon2024 \cite{wang2024cmrxrecon2024}. Both were acquired using a 3T scanner (MAGNETOM Vida, Siemens Healthineers, Germany) with a dedicated 32-channel cardiac coil, which was compressed into ten virtual coils. Participants were scanned in a prone position, with the cardiac cycle segmented into 12-25 phases, achieving a temporal resolution of approximately 50 ms, depending on the heart rate. Both datasets feature a spatial resolution of 1.5 × 1.5 mm² and a slice thickness of 8 mm. The CMRxRecon2023 includes cine imaging using the TrueFISP sequence and T1/T2 mapping with MOLLI-FLASH and T2prep-FLASH sequences, respectively, across various anatomical views such as short-axis and long-axis (2, 3, and 4 chambers). This dataset consists of a training set with 120 fully sampled subjects and uses 4x, 8x, and 10x uniform acceleration masks for under-sampling the complete k-spaces. The CMRxRecon2024 expands upon this with additional contrasts and views, including tagging, aorta, flow2D, and BlackBlood contrasts, incorporating diverse anatomical perspectives such as the left ventricular outflow tract (LVOT) for the cine contrast and aortic views (transversal and sagittal). Flow2D and BlackBlood contrast images are used exclusively for validation as unseen data. This dataset contains a larger training set with 200 fully sampled subjects. It provides a variety of k-space under-sampling trajectories (uniform, Gaussian, pseudo-radial) with multiple acceleration factors (4x, 8x, 12x, 16x, 20x, 24x) than the CMRxRecon2023 under-sampling trajectories.

\subsection{The Proposed Approach}
The optimization problem of equation \eqref{eq1} must be solved based on the compressed sensing theory \cite{donoho2006compressed} to obtain the high-quality clinical interpretable image \(x\) from the highly subsampled multi-coil k-space \(y\).

\begin{equation}
\min_{x} \frac{1}{2} \lVert y-Ax \rVert^{2}_{2} + \lambda R(x)
\label{eq1}
\end{equation}

\noindent where \(A\) and \(R(x)\) are measurement matrix and sparsity regularization term of \(x\), respectively. \(\lambda\) is used as a weighting coefficient for the \(R(x)\).

This optimization problem has been solved iteratively by \cite{sriram2020end} in the form of equation \eqref{eq2}.

\begin{equation}
k^{(t+1)} \leftarrow k^{(t)} - \eta^{(t)} * \mathscr{M} * (k^{(t)} - k^{(0)}) + G_k
\label{eq2}
\end{equation}

\noindent where in an iterative quality improvement process, \(k^{(t)}\) and \(k^{(t+1)}\) are the k-spaces of \(t\) and \(t+1\) stages. So, \(k^{(0)}\) is the initial adjacent multi-coil subsampled k-spaces \([..., k^{(t)}_{centeral-1}, k^{(t)}_{centeral}, k^{(t)}_{centeral+1}, ...] \) at \(t=0\). These adjacent k-spaces are similar slices in adjacent cardiac phases/weightings of \(k^{(t)}_{centeral}\). \(\eta^{(t)}\) and \(\mathscr{M}\) are also learnable step sizes and subsampling masks, respectively. \(G_k\), the gradient of the sparsity regularization term \(R(x)\), is obtained in a sequence of processes to complete the above equation (See Algorithm \ref{alg1}).

\begin{figure}[!t]
\centerline{\includegraphics[width=\textwidth]{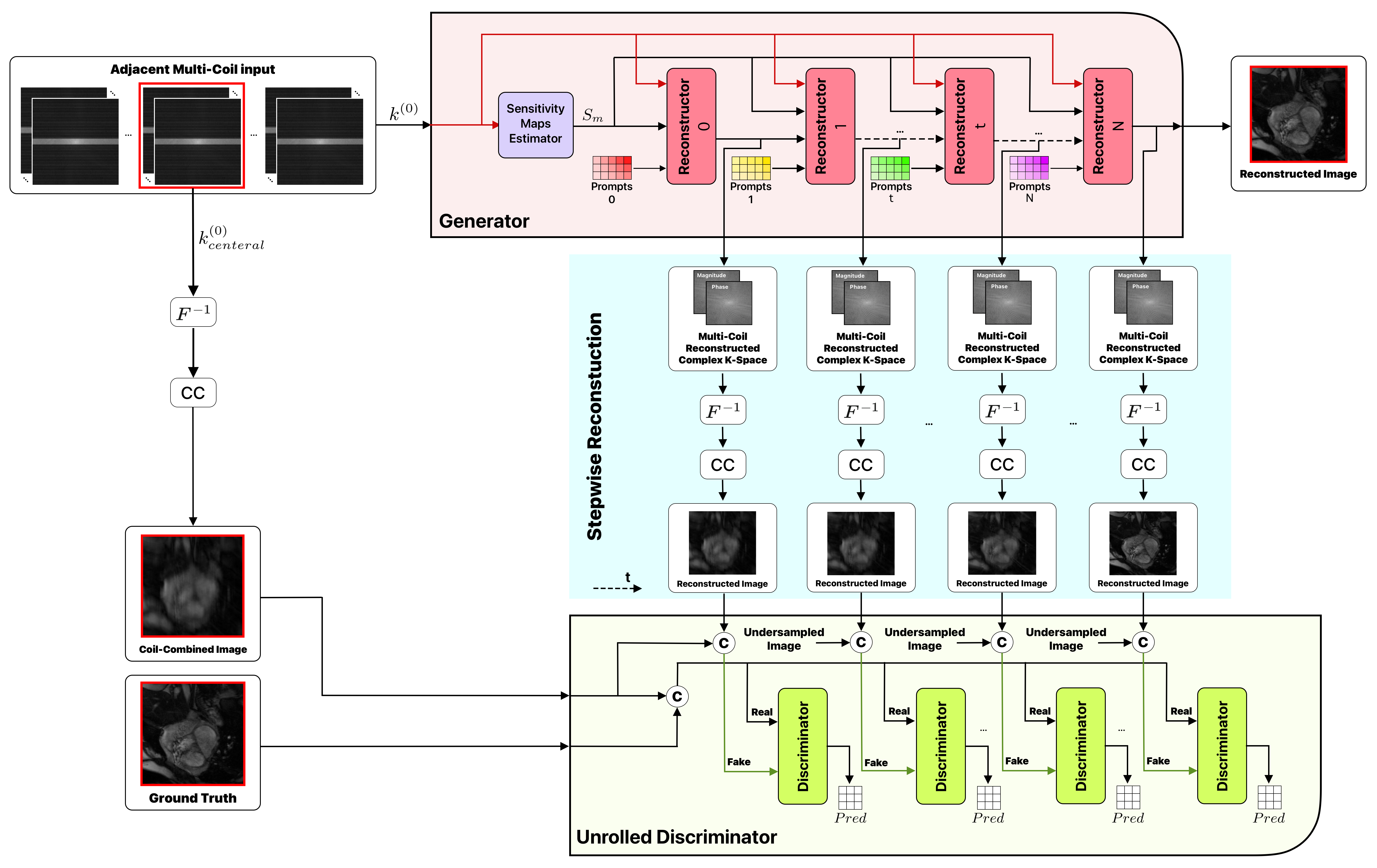}}
\caption{The overview of the proposed approach.}
\label{fig1}
\end{figure}

Figure \ref{fig1} shows the overview of the proposed approach. This approach is proposed based on the Patch-GAN architecture \cite{isola2017image} which ensures that regions of the reconstructed image are indistinguishable from the corresponding ones in the ground-truth image. \(k^{(0)}\) is received from the input, and in a stepwise reconstruction process, a high-quality image is obtained. In the stepwise reconstruction process, \(k^{(t+1)}\) is generated through equation \eqref{eq2} as an output of reconstructor \(t\). In Figure \ref{fig1}, each reconstructor module performs a sequence of operations to reconstruct \(k^{(t+1)}\), in part of which an attention-based prompt UNet (APUNet) including prompt blocks inspired by \cite{potlapalli2024promptir, xin2023fill} is used. A fixed prompt is a vector of fixed length and random values. Each APUNet receives fixed prompts as many as its levels. At each level of APUNet, a fixed prompt and features obtained from a lower-level block are entered into the discriminative prompt block module. This module generates an input-type adaptive prompt that helps APUNet adjust the reconstruction process to the input type at each level. Upon receiving \(k^{(0)}\), \(k^{(t)}\), conjugate symmetric sensitivity map \(S^{'}_{m}\), and prompt \(P^{(t)}\), each reconstructor module tries to reconstruct \(k^{(t+1)}\). From the output of each reconstructor module, a physical loss \(\mathscr{L}^{(t)}_{Phys}\) is calculated in the frequency space, including the sum of the mean squared errors (MSE) of the magnitude (\(Mag\)) and phase (\(\Phi\)) of \(k^{(t+1)}_{centeral}\) and the ground-truth k-space \(k^{(G)}\). \(k^{(t+1)}_{centeral}\) represents the central k-space, which the reconstruction process is performed to refine, without considering its adjacent k-spaces. Besides, the loss function \(\mathscr{L}^{(t)}_{SSIM}\) is computed using the structural similarity index measure (SSIM) \cite{wang2004image} by transforming \(k^{(t+1)}_{centeral}\) and \(k^{(G)}\) to the image space using the inverse fast Fourier transform (FFT) and then coil combination (CC), which is added to the physical loss \(\mathscr{L}^{(t)}_{Phys}\) to form stepwise loss \(\mathscr{L}^{(t)}_{step}\).

\begin{figure}[!t]
\centerline{\includegraphics[width=\textwidth]{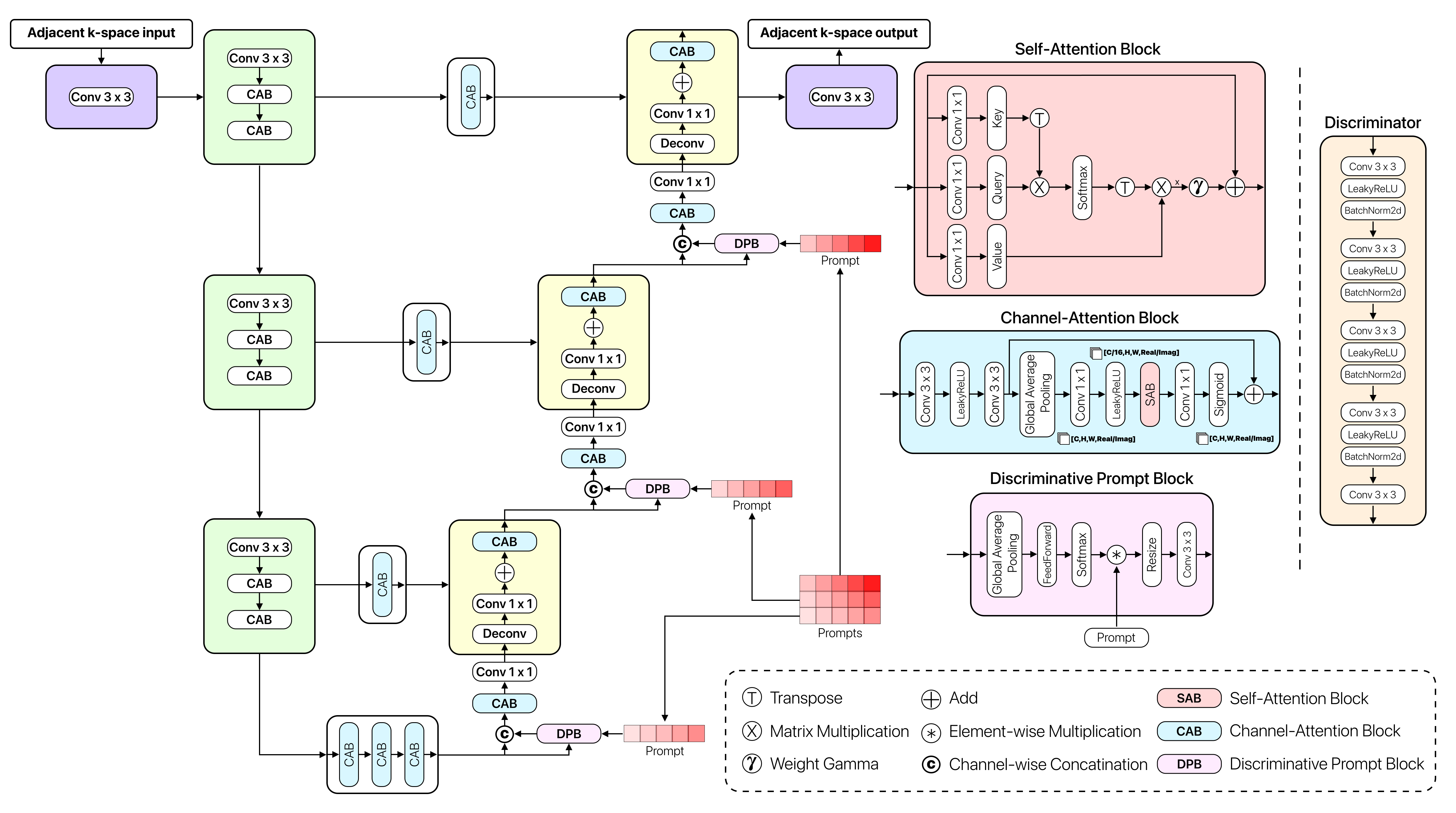}}
\caption{The structures of the APUNet and the discriminator networks.}
\label{fig2}
\end{figure}

Zero-filled image concatenation with the output of each reconstructor and the ground-truth image to create \(Input_{real}\) and \(Input_{fake}\) causes the discriminator to learn the distortions resulting from subsampling. This issue can help the generator to achieve more efficient performance in the face of unseen data. Given \(Input_{real}\) and \(Input_{fake}\), the discriminator network predicts \(Pred_{real}\) and \(Pred_{fake}\), then uses binary cross-entropy (BCE) \cite{reid2010composite} loss to approximate the discriminator loss \(\mathscr{L}_{Disc}\), which is used to learn the discriminator network. This process is repeated \(t\) times until the final \(k^{(t+1)}\) is achieved. Using the output of each reconstructor module for the discriminator network training provides faster convergence. In the last stage, the generator loss \(\mathscr{L}_{GAN-Gen}\) should be calculated in the image space and added to the sum of the step losses \(\mathscr{L}^{(t)}_{step}\). The generator's loss function is estimated by equation \eqref{eq3}. This loss function is leveraged to learn the generator network, learnable step size \(\eta^{(t)}\), and the sensitivity map estimator network, which is an APUNet. The task of the sensitivity
map estimator is to receive the concatenated coil data to predict the sensitivity maps. This concatenation approach allows for flexibility in receiving data from the different number of coils. The APUNet and discriminator network structures are shown in Figure \ref{fig2}. The reason for using channel-attention blocks is the presence of adjacent k-spaces as input so that the features extracted from the adjacent k-spaces can be considered for the reconstruction of the central k-space. The training process of the proposed approach is summarized in Algorithm \ref{alg1}.

\begin{equation}
\mathscr{L}_{Gen} = \lambda * \sum_{i=1}^{N} \mathscr{L}^{(t)}_{step} + \mathscr{L}_{GAN-Gen}
\label{eq3}
\end{equation}

\noindent where \(\lambda\) indicates the stepwise loss adjustment coefficient.

\begin{algorithm*}
    \caption{The Proposed All-in-one Approach for CMRI Reconstruction }
    \label{alg1}
    \begin{algorithmic}[1]
      
      \REQUIRE 
      $ $
      \\
      $\triangleright\ k^{(0)}:$ Adjacent multi-coil subsampled k-spaces   
      \\
      $\triangleright\ k^{(G)}:$ Ground-truth multi-coil k-space
      \\
      $\triangleright\ P^{(t)}:$ Discriminative prompt 
      \\
      $\triangleright\ \mathscr{M}:$ Subsampling mask 
      \\
      $\triangleright\ \eta^{(t)}:$ Learnable step size
      \\
      $\triangleright\ Label_{real}:$ Zero-filled matrix
      \\
      $\triangleright\ Label_{fake}:$ One-filled matrix
      \\
      $\triangleright\ \lambda:$ Stepwise loss adjustment coefficient 
      \\
      \ENSURE 
      $ $
      \\
      $\triangleright\ \theta_{Gen}:$ Weights of the generator network 
      \newline
      \STATE $k^{(ACS)} \leftarrow \mathrm{Auto\_calibration\_signal} (k^{(0)}) $
      \STATE $I^{(ACS)} \leftarrow \mathrm{FFT^{-1}} (k^{(ACS)}) $
      \STATE $S_m  \leftarrow \mathrm{APUNet^{(SME)}} (I^{(ACS)}) $
      \STATE $S^{'}_{m} \leftarrow \mathrm{Conjugate\_Symmetry} (S_m)$
      \FOR{$iteration=1,2,...$}
      \FOR{$t=0,1,2,...$}
      \STATE  $I_{MC} \leftarrow \mathrm{FFT^{-1}} (k^{(t)}) $
      \STATE $I_{SC} \leftarrow \mathrm{Multiplication} (S^{'}_{m}, I_{MC}) $
      \STATE $I_{RF} \leftarrow \mathrm{APUNet^{(t)}} (I_{SC}, P^{(t)})$
      \STATE $I^{'}_{RF} \leftarrow \mathrm{Repeat\_interleaved} (I_{RF})$
      \STATE $I_{SS} \leftarrow \mathrm{Multiplication} (I^{'}_{RF},S_m )$
      \STATE $G_k \leftarrow \mathrm{FFT} (I_{SS})$
      \STATE $k^{(t+1)} \leftarrow k^{(t)} - \eta^{(t)} * \mathscr{M} * (k^{(t)} - k^{(0)}) + G_k $
      \STATE $\mathscr{L}^{(t)}_{Phys} \leftarrow \mathrm{MSE} (Mag(k^{(t+1)}_{centeral}), Mag(k^{(G)})) + \mathrm{MSE} (\Phi(k^{(t+1)}_{centeral}), \Phi(k^{(G)}))$
      \STATE $\mathscr{L}^{(t)}_{SSIM} \leftarrow \mathrm{SSIM\_Loss} (CC(FFT^{-1}(k^{(t+1)}_{centeral})), CC(FFT^{-1}(k^{(G)}))) $
      \STATE $\mathscr{L}^{(t)}_{step}  \leftarrow \mathscr{L}^{(t)}_{Phys} + \mathscr{L}^{(t)}_{SSIM}$
      \STATE $Input_{real} \leftarrow \mathrm{Concat} (CC(FFT^{-1}(k^{(G)})), CC(FFT^{-1}(k^{(0)}_{centeral})))$
      \STATE $Input_{fake} \leftarrow \mathrm{Concat} (CC(FFT^{-1}(k^{(t+1)}_{centeral})), CC(FFT^{-1}(k^{(0)}_{centeral})))$
      \STATE $Pred_{real}, Pred_{fake}  \leftarrow \mathrm{Discriminator} (Input_{real}), \mathrm{Discriminator} (Input_{fake}) $
      \STATE $\mathscr{L}_{real} \leftarrow \mathrm{BCE\_Loss} (Label_{real}, Pred_{real})$
      \STATE $\mathscr{L}_{fake} \leftarrow \mathrm{BCE\_Loss} (Label_{fake}, Pred_{fake})$
      \STATE $\mathscr{L}_{Disc} \leftarrow \frac{1}{2} (\mathscr{L}_{real} + \mathscr{L}_{fake})$
      \STATE $\theta_{Disc} \leftarrow$ Adamw\_Optimizer $(\mathscr{L}_{Disc}, \theta_{Disc}) $
      \ENDFOR 
      \STATE $Input_{fake} \leftarrow \mathrm{Concat} (CC(FFT^{-1}(k^{(t+1)}_{centeral})), CC(FFT^{-1}(k^{(0)}_{centeral})))$
      \STATE $Pred_{fake}  \leftarrow \mathrm{Discriminator} (Input_{fake}) $
      \STATE $\mathscr{L}_{GAN-Gen} \leftarrow \mathrm{BCE\_Loss} (Label_{real}, Pred_{fake})  $      
      \STATE $\mathscr{L}_{Gen} \leftarrow \lambda * \sum_{i=1}^{N} \mathscr{L}^{(t)}_{step} + \mathscr{L}_{GAN-Gen}$
      \STATE $\theta_{Gen},\theta_{SME} \leftarrow$ Adamw\_Optimizer $([\mathscr{L}_{Gen}], \theta_{Gen},\theta_{SME}) $
      \ENDFOR 
      \RETURN $\theta_{Gen} \hspace*{8.1em}  \mathrm{\triangleright\ Weights\ of\ the\ generator\ network}$
    \end{algorithmic}

\end{algorithm*}

In our approach to reconstructing undersampled k-spaces using a stepwise Patch-GAN architecture, specific hyperparameters were employed in both the generator and discriminator networks. We deployed the AdamW optimizer \cite{loshchilov2017decoupled} with a learning rate of 0.002, weight decay of 0.1, and a gradient clipping value of 0.1. The learning rate was scheduled to be adjusted by a scheduler with a step size of 11 and a gamma of 0.1. The generator network consists of 12 reconstructor modules, and we used 16 auto-calibration lines for the undersampled k-spaces with an adjacent k-space length of 5. Curriculum learning \cite{bengio2009curriculum} starts with lower acceleration factors, allowing the model to grasp the fundamentals of reconstruction in an easy-to-hard manner. As the model progressively learned to handle these tasks, we used transfer learning to fine-tune it with higher acceleration factors. This method enabled the model to build on its initial understanding and adapt to more complex scenarios, gradually increasing the complexity of the task and improving overall performance and efficiency. The training was done in 12 epochs with a batch size of 1. Besides, the performance of our proposed method is evaluated using the SSIM, peak signal-to-noise ratio (PSNR), and normalized MSE (NMSE) metrics \cite{wang2004image}. The model has about 100 million parameters, so to support the computational requirements of the proposed model, 2x NVIDIA GPU H100 were used for training and evaluations.

\section{Experimental Results and Discussion}
\begin{figure}[!t]
\centerline{\includegraphics[width=\textwidth]{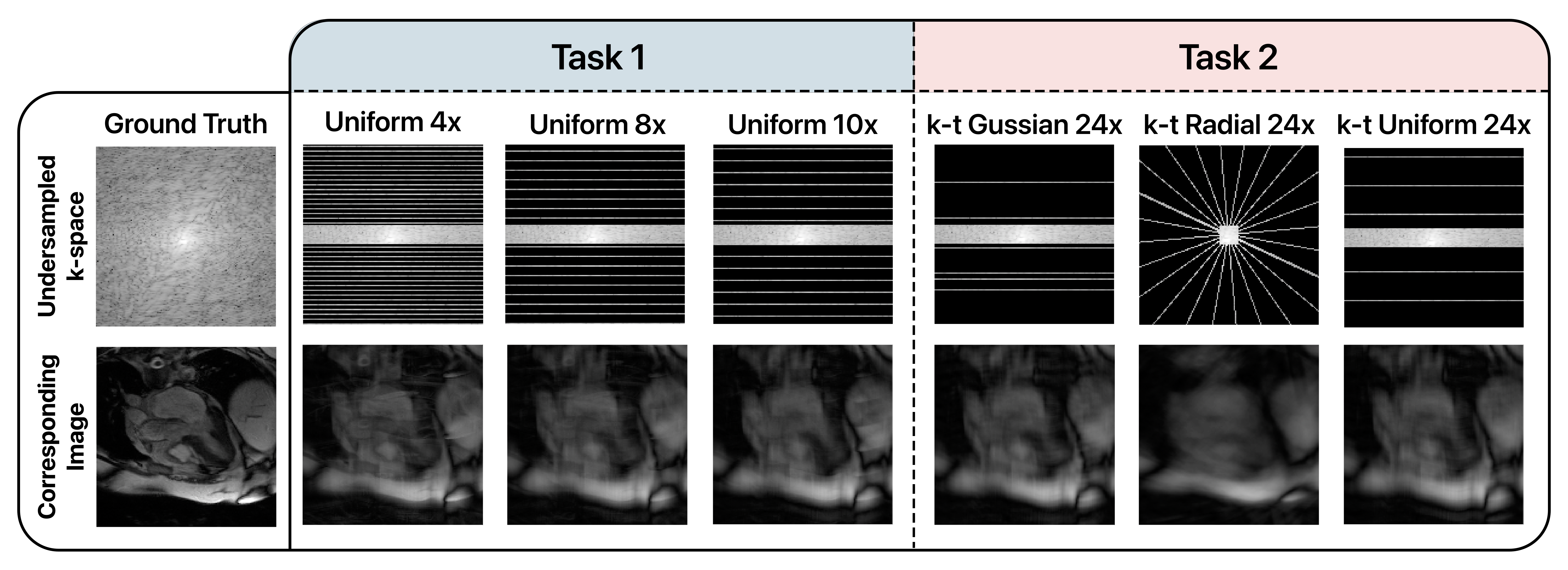}}
\caption{An example of several k-space subsampling masks and corresponding images.}
\label{fig3}
\end{figure}

In this section, the evaluation results of the proposed approach are presented with extensive experiments, and then the role of each component in the approach is analyzed by ablation studies. An example of several k-space subsampling patterns of the CMRxRecon2024 challenge is shown in Figure \ref{fig3}. In this figure, the result of k-space masking using the masks introduced is depicted in the image space after the inverse Fourier transform.

\begin{figure}[!t]
\centerline{\includegraphics[width=\textwidth]{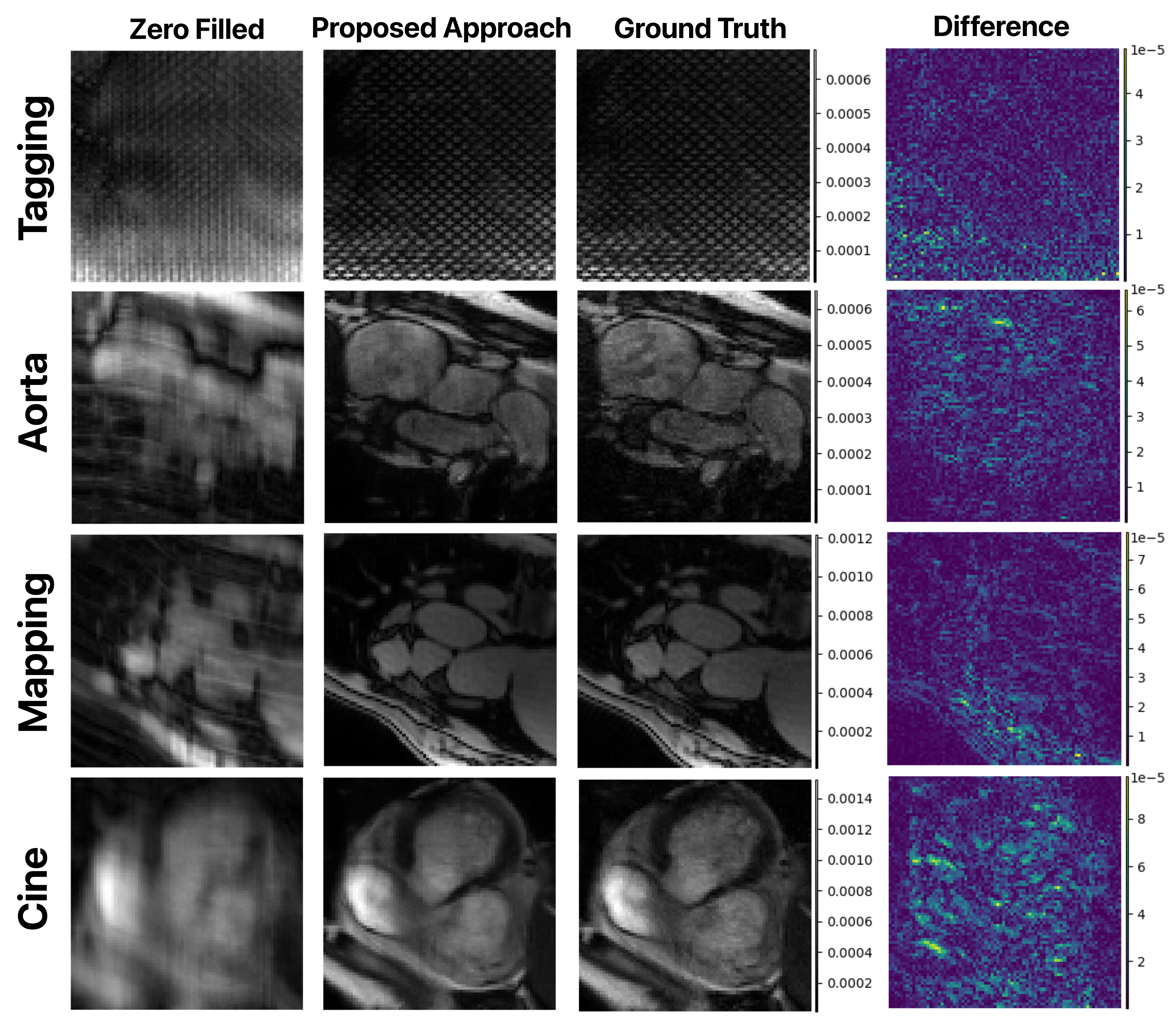}}
\caption{An example of qualitative comparison of images reconstructed from subsampled k-space using a uniform sampling pattern with a 10x acceleration factor, reconstructed by the proposed approach and ground truth images.}
\label{fig4}
\end{figure}

The experimental results related to evaluating the proposed approach in the two tasks introduced by the CMRxRecon2024 challenge are shown in Table \ref{tab1}. In Task 1, three different models for three acceleration factors of 4, 8, and 10x of uniform subsampling have been trained using the curriculum learning method. The average results of the three trained models are presented separately for different contrasts based on NMSE, PSNR and SSIM metrics. In task 2, a model has been trained using k-space data with various trajectories and subsampling acceleration factors of 4 to 24x using the curriculum learning method. Besides, the results of the proposed approach have been compared with two image reconstruction methods and the state-of-the-art PromptMR \cite{xin2023fill} method, which is the first rank of the CMRxRecon2023 challenge. An example of a qualitative comparison of images reconstructed from subsampled k-space, reconstructed by the proposed approach and ground truth images is shown in Figure \ref{fig4}.

\begin{table*}[!htp]\centering
\caption{The results of the proposed approach versus various MRI reconstruction methods and the state-of-the-art one for the first and second tasks of the CMRxRecon2024 challenge. The results are reported based on NMSE/PSNR/SSIM metrics, and the best values are shown in boldface.}\label{tab1}
\resizebox{\textwidth}{!}{
\renewcommand{\arraystretch}{1.5}
\begin{tabular}{c|c|c|c|c|c|c|c|c|c}\toprule
 \multirow{3}{*}{\textbf{Task}} &  \multirow{3}{*}{\textbf{Method}} &\multicolumn{3}{c|}{\textbf{Cine}}  &\multicolumn{2}{c|}{\textbf{Mapping}} &\multicolumn{2}{c|}{\textbf{Aorta}} &\textbf{Tagging} \\\cmidrule(lr){3-10}
& &SAX &LAX &LVOT &T1W &T2W &TRA &SAG &SAX \\\cmidrule(lr){1-10}
\multirow{4}{*}{1} &PromptIR \cite{potlapalli2024promptir} &2.6/40.05/0.9643 &2.8/38.42/0.9567 &2.9/38.12/0.9513 &2.4/38.87/0.9612 &1.6/42.38/0.9778 &2.0/41.21/0.9689 &2.2/40.61/0.9743 &1.9/41.41/0.9786 \\\cmidrule(lr){2-10}
&E2E-Varnet \cite{sriram2020end} &1.7/41.95/0.9723 &2.3/39.23/0.9632 &2.1/39.63/0.9648 &1.7/42.27/0.9733 &1.6/42.32/0.9763 &1.9/41.62/0.9754 &2.0/41.59/0.9739 &1.5/42.56/0.9791 \\\cmidrule(lr){2-10}
&PromptMR \cite{xin2023fill} &1.1/45.34/0.9821 &1.2/43.27/0.9814 &1.2/43.32/0.9829 &1.1/45.23/0.9806 &0.8/46.15/0.9878 &0.9/45.67/0.9873 &1.1/45.12/0.9813 &0.7/46.34/0.9912 \\\cmidrule(lr){2-10}
&Proposed Approach &\textbf{0.8/47.13/0.9903} &\textbf{0.9/46.12/0.9893} &\textbf{0.9/46.40/0.9901} &\textbf{0.6/48.34/0.9897} &\textbf{0.6/48.50/0.9919} &\textbf{0.7/47.54/0.9914} &\textbf{0.8/46.81/0.9897} &\textbf{0.5/49.80/0.9931} \\\midrule
\multirow{4}{*}{2} &PromptIR \cite{potlapalli2024promptir} &3.2/37.95/0.9468 &3.3/37.12/0.9351 &3.4/37.00/0.9300 &3.0/36.87/0.9505 &2.5/40.32/0.9578 &2.8/39.62/0.9589 &3.0/38.42/0.9545 &2.5/40.41/0.9686 \\\cmidrule(lr){2-10}
&E2E-Varnet \cite{sriram2020end} &2.7/38.42/0.9531 &3.0/38.29/0.9472 &2.9/38.63/0.9450 &2.6/39.27/0.9533 &2.2/41.38/0.9653 &2.6/40.21/0.9654 &2.8/39.61/0.9643 &2.5/41.56/0.9691 \\\cmidrule(lr){2-10}
&PromptMR \cite{xin2023fill} &2.1/40.13/0.9649 &2.2/41.25/0.9654 &2.2/41.32/0.9629 &2.1/42.23/0.9606 &1.5/44.15/0.9778 &1.6/44.21/0.9773 &1.8/43.12/0.9713 &1.4/45.34/0.9812 \\\cmidrule(lr){2-10}
&Proposed Approach &\textbf{1.4/42.52/0.9774} &\textbf{1.4/42.43/0.9782} &\textbf{1.3/43.40/0.9781} &\textbf{1.0/45.34/0.9717} &\textbf{1.0/46.50/0.9829} &\textbf{1.1/45.54/0.9824} &\textbf{1.2/44.81/0.9797} &\textbf{0.9/48.80/0.9891} \\
\bottomrule
\end{tabular}}
\end{table*}

Comparing our approach with the E2E-Varnet and PromptMR methods, a notable distinction lies in the strategies employed for image reconstruction. E2E-Varnet integrates the learning of sensitivity maps during model training and k-space intermediate quantities utilization for reconstruction. This approach has shown superior performance in estimating sensitivity maps when presented with limited sample lines for auto-calibration signals. PromptMR extends the E2E-Varnet method using the discriminative prompt block module for handling various input types. It further refines reconstructed images utilizing adjacent inputs and employing a video restoration model to enhance image quality, albeit at the cost of added complexity and a departure from a singular model architecture. In contrast, our approach leverages a Patch-GAN architecture for CMR image reconstruction, emphasizing parity between reconstructed and ground truth images through GAN architecture. By adopting a stepwise reconstruction method and incorporating two distinct physical and spatial loss functions at each reconstruction step, we address the issue of the vanishing gradient. Compared to the PromptMR, our method does not require a quality improvement step through a video restoration model and performs the reconstruction process in an integrated manner.

Several experiments were conducted to evaluate the role of different components in the proposed approach. These ablation studies determined the role of self-attention blocks, physical loss function, stepwise loss calculations, curriculum learning and discriminator network. These experiments were performed on cine CMR images of the first task of the CMRxRecon2024 challenge. The results of these ablation studies are given in Table \ref{tab2}.

\begin{table}[!htp]\centering
\caption{The results of ablation studies to investigate the role of different components of the proposed approach. The results are reported based on NMSE/PSNR/SSIM metrics, and the best values are shown in boldface.}\label{tab2}
\resizebox{0.8\textwidth}{!}{
\begin{tabular}{l|c|c|c|c}\toprule
\multirow{2}{*}{} &\multicolumn{3}{c|}{\textbf{Cine}} \\\cmidrule(lr){2-4}
&SAX &LAX &LVOT \\\midrule
Excluding Self-Attention Blocks &1.1/45.69/0.9795 &1.2/45.21/0.9803 &1.2/45.01/0.9843 \\\midrule
Without Physical Loss &0.9/46.54/0.9892 &1.0/45.82/0.9848 &0.9/46.06/0.9861 \\\midrule
Without Stepwise Loss &0.9/46.47/0.9878 &1.0/45.71/0.9832 &1.0/45.99/0.9848 \\\midrule
Without Curriculum Learning &0.9/46.92/0.9887 &1.0/45.98/0.9845 &1.0/46.11/0.9862 \\\midrule
Generator Only &1.0/46.52/0.9885 &1.1/45.89/0.9834 &1.1/46.05/0.9862 \\\midrule
The Proposed Approach &\textbf{0.8/47.13/0.9903} &\textbf{0.9/46.12/0.9893} &\textbf{0.9/46.40/0.9901} \\
\bottomrule
\end{tabular}
}
\end{table}

By simultaneously considering the various acceleration factors, the multi-contrast characteristics of CMR imaging, the diverse anatomical views, and different k-space sampling trajectories, our all-in-one approach represents a holistic solution for accelerating CMR imaging. This approach facilitates the generation of high-quality images, ensures more accurate diagnoses, and enhances the overall patient experience.

\section{Conclusion}
A universal approach for multi-contrast CMR image reconstruction was proposed in this study. Based on the Patch-GAN architecture, the proposed all-in-one approach receives subsampled multi-contrast k-space with different acceleration factors and trajectories and generates a high-quality clinical interpretable image. In this image reconstruction process, the advantages of k-space and image domain features are simultaneously used through physical and SSIM loss functions. The image refinement is based on a stepwise process, and the stepwise loss function is defined to avoid the vanishing gradient problem, which is finally added as a term to the generator loss function. Besides, the curriculum learning was leveraged in the proposed approach training, so the learning process from simple to hard tasks improved the overall performance. Evaluation of the proposed approach with extensive experiments and ablation studies showed that this approach outperforms previous methods. The future plan will be to improve the model's generality in the domain shift conditions between input datasets by relying on domain adaptation methods.

%
%
%

\bibliographystyle{splncs04}


\end{document}